\documentstyle[aps,preprint]{revtex}
\begin{document}
\draft
\preprint{IC/94/z; IMSc-94/46}
\title{Shock Wave Mixing in Einstein and Dilaton Gravity}
\author{Saurya Das \footnote{E-Mail:~saurya@imsc.ernet.in}}
\address{The Institute of Mathematical Sciences, \\ CIT Campus,
Madras - 600 113,  India.}
\author{Parthasarathi Majumdar\footnote{E-Mail:~partha@ictp.trieste.it;
partha@imsc.ernet.in}}
\address{International Centre for Theoretical Physics, Trieste I-34100,
Italy\footnote{Permanent address: The Institute of Mathematical Sciences,
Madras 600 113, India.}}
\maketitle
\begin{abstract}
We consider possible mixing of electromagnetic and gravitational shock
waves, in the Planckian energy
scattering of point particles in Minkowski space. By
boosting a Reissner-Nordstr\"om black hole solution to the velocity of
light, it is shown that no mixing of shock waves
takes place for arbitrary finite charge carried by the black hole.
However, a similar boosting procedure for a charged black hole
solution in dilaton gravity yields some mixing : the wave function of even
a neutral test particle,
acquires a small additional phase factor depending on the dilatonic
black hole charge. Possible implications for poles in the amplitudes for
the dilaton gravity case are discussed.
\end{abstract}

\section{Introduction}

The predominance of shock waves as instantaneous mediators of gauge and
gravitational interactions in two particle scattering at Planckian centre
of mass energies and fixed low momentum transfers has been the subject of
some interest recently
\cite{thf},\cite{ver1},\cite{jac},\cite{ver2},\cite{dm}. All
purely local gauge and/or gravitational field degrees of freedom decouple
in this kinematical regime. The residual degrees of freedom are basically
gauge or coordinate transformation parameters evaluated on the boundary
of the null plane, describing a simpler field theory whose classical
solutions represent appropriate shock waves. Two particle S-matrices are
exactly calculable from these reduced theories and reproduce, wherever
possible, the corresponding {\it eikonal} amplitudes
\cite{jac},\cite{amat}. There are of course situations when this
decoupling of local degrees of freedom is not easy to establish within a
field theory, as e.g., when magnetic monopoles
are present. However, this does not deter description of the
ensuing interactions in terms of shock waves, nor does it in any way
affect the computation of two particle amplitudes. The
interplay between electromagnetic and gravitational interactions is
particularly significant in this case since the magnetic monopole sector
is basically strongly coupled, akin to gravity at Planckian energies
\cite{dm}.

The gravitational shock wave relevant to two particle scattering in
Minkowski space has been obtained \cite{dray} in one of two ways: either
by demanding that the Minkowskian geometry with a
lightlike particle present is the same as an empty Minkowski space with
coordinates shifted along the geodesic of the particle, or by a process
of `boosting' the metric of a massive particle to luminal velocities when
its mass exponentially decays to zero.  Since there are well known
(electrically and magnetically) charged black hole solutions of the
Einstein equation, boosting such solutions to the velocity of light would
of course produce both gravitational and electromagnetic shock waves. It
is then important to determine whether these two species of shock waves
actually {\it mix}. The
problem may be stated succintly as follows: the calculation of the
amplitude in the shock wave picture
entails computing the phase factor that the shock wave induces on the
wave function of the target particle. When both particles carry charge,
phase factors are induced by electromagnetism and gravity
independently of the other. It has been assumed in the literature that ,
with both shock waves present, the net phase factor is simply the sum of
the individual phase factors \cite{thf},\cite{jac},\cite{dm}.
In other words, the gravitational and
electromagnetic shock waves are assumed to travel collinearly without
interaction,
even though they are extremely localized singular field configurations.
In our opinion, this assumption warrants justification. This is what
is attempted in the sequel.

The study of mixing of the two varieties of shock waves proceeds by
considering black hole solutions that also carry electric/magnetic charge,
and as such, produce both kinds of shock waves upon boosting a la'
\cite{dray}. In section II we show that a boosting
of the standard Reissner-Nordstr\"om black hole metric leads to
a decoupling of electromagnetic and gravitational effects in a somewhat
subtle manner: the part of the boosted metric that depends explicitly on the
charge,
can be removed by a diffeomorphism, leaving behind a piece which cannot,
being non-differentiable at $x^-=0$. This
`discontinuity' in the boosted metric, which constitutes the gravitational
shock wave, is identical to the discontinuity in the boosted Schwarschild
metric, independent of the charge of its parent black hole
solution. In the next section, we attempt an application of a similar
boosting procedure to charged
black hole solutions of four dimensional dilaton gravity. For generic
values of the charge, this endeavor results in a boosted metric
whose discontinuities do explicitly depend on the charge. More importantly,
in this case both null coordinates appear to warrant a discontinuous
transformation. We
discuss the implication of this in terms of singularities of the metric,
and show that the gravitational phase shift of the wave function of a test
particle
depends on the black hole charge even when the particle itself is
neutral. We end in section IV with a few concluding remarks.

\section{Decoupling in the Reissner-Nordstr\"om Case}

The gravitational field due to a stationary point particle of mass $M$ and
electric charge $Q$ is given by the standard Reissner-Nordstr\"om metric
\begin{equation}
ds^2~=~~(1-{2GM \over r} + {GQ^2 \over r^2}) dt^2
{}~-~(1-{2GM \over r} + {GQ^2 \over r^2})^{-1} dr^2 ~-~r^2 d \Omega^2~,
\label{rn}
\end{equation}
where $G$ is Newton's constant. The question we address here is : if the
particle is Lorentz-boosted to a velocity $\beta \sim 1$, what will be
the nature of the gravitational field as observed in the `stationary'
frame? Let us assume, for simplicity and without loss of generality, that
the particle is boosted in the $+z$ direction, so that $z,t$ are related
to the tranformed coordinates $Z,T$ according to
\begin{eqnarray}
T~&=&~t\cosh \rho~ + ~z\sinh \rho,     \nonumber \\
Z~&=&~t\sinh \rho~+~ z\cosh\rho~.
\label{eq:transf}
\end{eqnarray}
The parameter $\rho$ is called rapidity : $\beta=\tanh \rho$. The null
coordinates are defined as usual as $x^{\pm}=t \pm z$. The boosting
involves parametrizing the mass of the black hole as $M=2pe^{-\rho}$
where $p$ is the momentum of the boosted particle, lying almost entirely
in the longitudinal direction. $p$ is usually kept fixed at a large value
in the boosting process, and the limit of the boosted metric is evaluated
as $\rho \rightarrow \infty$. In this limit, the Reissner-Nordstr\"om
metric assumes the form
\begin{equation}
ds^2~\rightarrow~dx^- \{ dx^+~-~dx^- [{2Gp \over |x^-|}~-~{GQ^2 \over
(x^-)^2} ] \}~-~d x_{\perp}^2~~. \label{rnlim}
\end{equation}
The boosted metric does indeed seem to depend explicitly on the charge
$Q$. But, notice that this dependence is confined to a part of the
metric that can be removed by a diffeomorphism, albeit one that
is singular at the origin. However, the part that goes as $1/ |x^-|$
cannot be removed by any diffeomorphism; this latter, of course, is
precisely
the part that is associated with the gravitational shock wave \cite{thf}.
Further,
not only is its coefficient independent of $Q$, it is identical to the
the coefficient of the $1/ |x^-|$ term in the boosted Schwarschild
metric \cite{thf}, \cite{dray}. All
memory of the charge of the parent black hole solution is obliterated
upon boosting, insofar as the gravitational shock wave is concerned.
Consequently, the mutual transparency of the two shock waves follows
immediately. \footnote{We note here that our approach and results for the
Reissner-Nordstr\"om case differ somewhat from those of ref. \cite{san}
where, in fact, the limiting procedure employed appears to yield
vanishing electromagnetic shock waves in the luminal limit.} Hence the
net phase shift of the wave function of a test particle moving in the two
shock waves is simply the sum of the phase factors induced individually
by each shock wave. This, in the case of scattering of two electric
charges, simply amounts to the replacement $Gs \rightarrow Gs+ee'$ as
mentioned in ref.s \cite{thf}, \cite{jac}. A similar decoupling of
electromagnetic and gravitational shock waves can be seen by boosting a
{\it magnetically} charged Reissner-Nordstr\"om solution, which justifies
once again the determination of the net phase factor in the test charge
wave function as the sum of the individual phase factors in Planckian
charge-monopole scattering \cite{dm}.

The foregoing analysis is completely general, and requires no assumption
on the strength of the charge $Q$, except perhaps that it be finite.
However, an analysis of the singularities of the the metric in (\ref{rn})
indicates that, if one is to abide by the dictates of Cosmic Censorship,
the charge $Q$ must obey $Q \leq M$. In the extremal limit, the boosting
procedure adopted above forces $Q$ to decay exponentially to zero as
the rapidity runs to infinity. The electromagnetic shock wave, by
withering away in this limit, then trivially decouples from the
gravitational one. This was first pointed out in ref. \cite{san}.

\section{Non-decoupling in Dilaton Gravity}

The charged black hole solution of four dimensional dilaton gravity,
obtained as a part of an effective low energy theory from the heterotic
string compactified on some compact six-fold, is given, following
\cite{gib},\cite{gar}, as
\begin{equation}
ds^2~=~ (1-{\alpha \over Mr} )^{-1} \left [ (1 - {2GM \over r}) dt^2~-(1
-{2GM \over r})^{-1} dr^2~-~(1 -{\alpha \over Mr}) r^2 d \Omega^2
\right ]~~, \label{metr}
\end{equation}
where, $\alpha \equiv GQ^2 e^{-2\phi_0} $, with $\phi_0$ being the
asymptotic value of the dilaton field $\phi$. The metric reduces to the
Schwarschild metric when $\alpha=0$, and, not surprisingly, shares the
coordinate singularity at $r=2GM$ which becomes the event
horizon for the curvature singularity at $r=0$. In addition, there is
the `singularity' at $r={\alpha \over M}$ which is not necessarily a
coordinate singularity. We shall return to this point later.

We now apply the boosting procedure elaborated in the last section to
this metric. The mass of the black hole is parametrized as
$M=2pe^{-\rho}$ and the Lorentz-transformed metric is evaluated in the
limit as the rapidity $\rho \rightarrow \infty$ for fixed large $p$. The
result can be expressed as the Minkowski metric in terms of {\it shifted}
coordinate differentials,
$$ds^2~\rightarrow~d{\tilde x}^+ d{\tilde x}^-~-~(d{\tilde
x}_{\perp})^2~~, $$ where,
\begin{eqnarray}
d{\tilde x}^+ ~~&=&~~dx^+ ~-~\left ( { {4Gp \over |x^-|} \over {1 - {\alpha
\over p |x^-|}}} \right ) dx^-  \nonumber \\
d{\tilde x}^-~~&=&~~dx^- \left ( { {1 - {\alpha \over 2p |x^-|}} \over {1 -
{\alpha \over p |x^-|}} } \right )  \nonumber \\
d{\tilde {\vec x}}_{\perp}~~&=&~~d{\vec x}_{\perp}~~. \label{shft}
\end{eqnarray}
Several features emerge immediately from these equations; of these, the
most striking is the explicit dependence on the charge $\alpha$ of terms
that will surely contribute to the gravitational shock wave because of
their non-differentiable functional form. No less important is the fact
that, in this case, the coordinate $x^-$ which, in the Schwarschild (and
Reissner-Nordstr\"om) case(s), defined the null surface ($x^-=0$) along
which the two Minkowski spaces were to be glued, is now itself subject to
transformation by such a discontinuous function, again explicitly
depending on $\alpha$. Before examining these aspects in detail, we
note in passing that the results reduce to those in the Schwarschild case
in the limit $\alpha=0$, as indeed is expected.

First of all, the charge $\alpha$ may be chosen to be small by taking a
large value of $\phi_0$, so that, with a large value of $p$, one can
binomially expand the denominators in the rhs of the first two equations
in (\ref{shft}); this yields, for points away from $x^-=0$,
\begin{eqnarray}
d{\tilde x}^+~~&=&~~dx^+ ~-~{4Gp \over |x^-|} ~-~{4\alpha \over (x^-)^2}
{}~+~ {\cal O}(\alpha ^2 / p)~ \label{xpls} \\
d{\tilde x}^-~~&=&~~dx^-~+~{\alpha \over 2p |x^-|} ~+~ {\cal O}( \alpha^2 /
p^2)~. \label{xmns}
\end{eqnarray}
We now observe that, as far as the shift in $dx^+$ is concerned, the part
that will contribute to the gravitational shock wave is in fact, {\it
independent} of $\alpha$, and, furthermore, is identical to the result in
the Schwarschild and hence the Reissner-Nordstr\"om case. As for the latter
solution, the $\alpha$-dependent part may be rendered innocuous by a smooth
diffeomorphism.

This is however not the case for the shift in $dx^-$,
which is explicitly $\alpha$-dependent. Clearly, the gravitational shock
wave now possesses a more complicated geometrical structure than in the
earlier examples. The geometry can no longer be expressed as two
Minkowski spaces glued after a shift along the null surface $x^-=0$, for
now there is a {\it discontinuity} in the $x^-$ coordinate at that very
point, in contrast to the previous cases where it was continuous. This
discontinuity has a rather serious implication : unlike in the earlier
situation wherein the coordinate $x^-$ could well serve as the affine
parameter characterising the null geodesic of a test particle crossing
the gravitational shock wave (cf. \cite{dray}), a null geodesic is
actually {\it incomplete} in this situation. To see this in more detail,
consider the geodesic equations  of a very light particle moving in the
Lorentz-boosted metric (\ref{metr}),
\begin{eqnarray}
{\dot T}~~&=&~~\left({ {1- {\alpha \over Mr}} \over {1 - {2GM
\over r}} } \right ) E~~ \nonumber \\
r^2 {\dot \phi}~~&=&~~\left ( 1 - {2GM \over r} \right ) L~~ \nonumber \\
{\dot r}^2~~&=&~~(1- {\alpha \over Mr})^2 \left [ E^2~-~{L^2 \over r^2}
\left ({ {1 - {2GM \over r}} \over {1 - {\alpha \over Mr}} } \right )
\right ]~~.
\label{geod}
\end{eqnarray}
Unlike the geodesic equations for a boosted Schwarzschild metric, which
can be solved perturbatively in a power series in the mass $M$ (or
alternatively in the parameter $e^{-\rho}$ (where $\rho$ is the
rapidity),\cite{dray} these equations do not admit any perturbative
solution because of the singularity at $r = \alpha/M$. Taking recourse to
singular perturbation theory does not evade the problem; the definition
of a {\it continuous} affine parameter is not possible in this case. It
follows that the singularity in question must be a {\it curvature}
singularity. Although for the Reissner-Nordstr\"om case also, for generic
value of the charge, the singularity at $r=0$ is no longer hidden by the
event horizon, there are no other singularities away from this point. In
the present instance, the singularity at $r=0$ is actually protected by
the Schwarzschild horizon. One might consider imposing an extremal
condition on the
charge $\alpha$ (vid. \cite{gar}): $\alpha = M^2$ to mitigate the
circumstances. However, this limit is not interesting for our
purpose, for the same reason that the extremal Reissner-Nordstr\"om is not
-- the charge decays exponentially to zero with the rapidity going off to
infinity.

The non-decoupling of gravitational and electromagnetic effects that we
see here can be made more articulate if one proceeds to actually
calculate the phase shift of the wave function of a test particle
encountering the gravitational shock wave, notwithstanding the pathologies
delineated above. The equations (\ref{shft}) above for the
differentials are consistent with the following finite shifts, obtained
by generalizing results of \cite{dray},
\begin{eqnarray}
x^+ _{>}~~&=&~~x^+ _{<}~+ ~2Gp \ln \mu^2 r_{\perp}^2~~ \nonumber \\
x^-_{>}~~&=&~~x^-_{<}~+~{\alpha \over 2p} \ln \mu^2 r_{\perp}^2~~
\nonumber \\
{\vec x}_{>}~~&=&~~{\vec x}_{<}~~. \label{shftf}
\end{eqnarray}
With these, following \cite{thf} we can easily calculate the net phase
shift of the wave function of a test particle due purely to gravitational
effects:
\begin{equation}
 \Phi_{total}~~=~~(Gs~+~\alpha {k_{\perp}^2 \over 2s} ) \ln \mu^2
r_{\perp}^2~~. \label{phsgr}
\end{equation}
Here, $k_{\perp}$ is the transverse momentum of the test particle.
Thus, even if the test particle is electromagnetically neutral, its wave
function undergoes a phase shift that depends on the charge of the black
hole boosted to produce the gravitational shock wave. This is a novel
phenomenon, in our opinion, although, strictly speaking, in the
kinematical regime under consideration, the magnitude of the effect is
small. Nevertheless, the mixing of the electromagnetic and gravitational
shock waves, in this case is quite obvious.

The scattering amplitude for a test particle encountering such a
gravitational shock wave can be calculated following ref. \cite{thf}.
Modulo standard kinematical factors and irrelevant constants, the answer
is
\begin{equation}
f(s,t)~~\sim~~{1 \over t} { {\Gamma \left(1 - i(Gs + \alpha {k_{\perp}^2
\over 2s} ) \right )} \over {\Gamma \left(i (Gs + \alpha {k_{\perp}^2
\over 2s}) \right ) } }~~. \label{ampl}
\end{equation}
Since the calculation is performed in a coordinate frame in which the
test particle is assumed to be moving slowly, the amplitude does not
appear manifestly Lorentz-invariant, although there is nowhere any
violation of Lorentz symmetry. A more refined calculation where Lorentz
invariance is explicitly maintained can indeed be done following Verlinde
and Verlinde\cite{ver1}, but will not be reported here. The only likely
outcome of such a calculation will be the replacement of the quantity
$k_{\perp}^2$ by the squared momentum transfer $t$ upto some numerical
coefficient of ${\cal O}(1)$. As a consequence, the poles in (\ref{ampl})
would undergo a shift of ${\cal O}(i \alpha Gt/N^2)$ from their
integer-valued (given by $N$) positions on the imaginary axis found in
the Schwarzschild case \cite{thf}. This shift is quite different from
similar shifts when electromagnetic effects are included based upon a
decoupling assumption \cite{thf}, \cite{jac}, \cite{dm}. The
non-decoupling is manifest from the coefficient $\alpha G$ in this case.
Also, the electromagnetic shifts are always constant independent of $t$,
in contrast to what we find here.

\section{Conclusions}

The decoupling of electromagnetic and gravitational shock waves have now
been established for the case of general relativity, justifying thereby
earlier results incorporating both fields for electrically and
magnetically charged particles scattering at Planckian centre-of-mass
energies. The shifts in the poles due to electromagnetic effects stand
vindicated. Admittedly, it is true that in
the Einstein gravity case the poles appear to be artifacts of the large
impact parameter approximation \cite{ver1}. But, the nature of the shift
due to the dilaton coupling tends to reinforce the speculation
that string theory may actually provide a way to compute
corrections to this approximation as a power series in $t$. It would be
interesting if
this behaviour could be retrieved from the high energy string amplitudes
calculated in earlier work \cite{amat} in some suitable local field
theory limit.

The results may also have implications for black holes. The effect of
infalling
particles collapsing gravitationally onto a black hole has been analyzed
\cite{dray} to produce a shift of the classical event horizon. If we also
subscribe to the view \cite{thfbh} that this shift essentially involves
generalizing the flat space gravitational shock wave to a curved background,
then a particle whose fields are obtained by boosting fields of a dilaton
black hole would cause extra shifts of the horizon of a
Schwarzschild black hole. In addition, with electric and magnetic
charges present, novel contributions are to be expected for any
S-matrix  (a la' 't Hooft \cite{thfbh}) one may attempt to propose for
dilatonic black holes. We hope to report on these in the near future.

We thank T. Jayaraman, R. Shankar and K. Subramaniam for useful
discussions. One of us (P.M.) also thanks F. Hussein, S. Mukherjee, M. A.
Namazie, K. Narain
and S. Randjbar-Daemi for very fruitful discussions, and the
International
Centre for Theoretical Physics, Trieste, Italy for excellent hospitality
during an Associateship visit during which this work was completed.

\end{document}